\documentclass[a4paper, extra, camera,%
]{gji}
\usepackage{timet}

\usepackage[utf8]{inputenc}
\usepackage{lmodern}
\usepackage[T1]{fontenc}
\usepackage{amssymb, amsmath, amsfonts}
\usepackage{booktabs}
\usepackage{xspace}
\usepackage[final]{graphicx}
\usepackage[position=top]{subfig}
\usepackage{ifthen}
\usepackage[pdftex, final, allcolors=blue, colorlinks=true]{hyperref}

\makeatletter
\let\zz@tabular\@tabular
\let\zzendtabular\endtabular
\let\zz@xtabularcr\@xtabularcr
\let\zz@tabclassz\@tabclassz
\let\zz@tabclassiv \@tabclassiv 
\let\zz@tabarray\@tabarray
\makeatother

\newlength{\fwidth}
\newlength{\cwidth}
\newlength{\twidth}
\ifthenelse{\boolean{GJ@referee}}{%
  \setlength{\cwidth}{234pt}%
  \setlength{\fwidth}{468pt}%
  \setlength{\twidth}{468pt}}{%
  \setlength{\fwidth}{504pt}%
  \ifthenelse{\boolean{@twocolumn}}{%
    \setlength{\cwidth}{240pt}%
    \setlength{\twidth}{240pt}}{%
    \setlength{\cwidth}{252pt}%
    \setlength{\twidth}{504pt}%
  }
}
\graphicspath{{./figures/}}

\usepackage{siunitx}
\usepackage{tabularx}
\usepackage{tcolorbox}

\AtBeginDocument{
  \label{CorrectFirstPageLabel}
  
}

\newcommand{\mr}[1]{\mathrm{#1}}
\newcommand{\emg}[2]{\texttt{emg#1#2}\xspace}
\newcommand{\empymod}{\texttt{empymod}\xspace}
\newcommand{\simpeg}{\texttt{SimPEG}\xspace}
\newcommand{\discretize}{\texttt{discretize}\xspace}
\newcommand{\custem}{\texttt{custEM}\xspace}
\newcommand{\petgem}{\texttt{PETGEM}\xspace}

\newcommand{\ohmm}{\ensuremath{\Omega\,}\text{m}\xspace}

\title[3D CSEM Open-Source Landscape]{Towards an open-source landscape for 3D CSEM modelling}

\author[D. Werthmüller \emph{et al.}]
  {\Large Dieter Werthmüller$^1$,  
   Raphael Rochlitz$^2$,           
   Octavio Castillo-Reyes$^3$, and 
   Lindsey Heagy$^4$\\             
   \footnotesize
  $^1$ TU Delft, Building 23, Stevinweg 1 / PO-box 5048, 2628 CN Delft (NL).
  E-mail: \href{mailto:Dieter@Werthmuller.org}{Dieter@Werthmuller.org}\\[-.3em]
   \footnotesize
  $^2$ Leibniz Institute for Applied Geophysics, Stilleweg 2, 30655 Hannover (DE)\\[-.3em]
   \footnotesize
   $^3$ Barcelona Supercomputing Center (BSC), Nexus II Building c/Jordi Girona, 29, 08034 Barcelona (ES)\\[-.3em]
   \footnotesize
  $^4$ University of California Berkeley, Department of Statistics (USA)
  }

\date{Received 2021 MMMM DD; in original form 2020 October 23}
\pagerange{\pageref{firstpage}--\pageref{lastpage}}
\volume{VVV}
\pubyear{202Y}

\begin{document}

\label{firstpage}

{\makeatletter
\let\@tabular\zz@tabular
\let\endtabular\zzendtabular
\let\@xtabularcr\zz@xtabularcr
\let\@tabclassz\zz@tabclassz
\let\@tabclassiv \zz@tabclassiv 
\let\@tabarray\zz@tabarray
\maketitle
}

\begin{summary}
Large-scale modelling of three-dimensional controlled-source electromagnetic (CSEM) surveys used to be feasible only for large companies and research consortia. This has changed over the last few years, and today there exists a selection of different open-source codes available to everyone. Using four different codes in the Python ecosystem, we perform simulations for increasingly complex models in a shallow marine setting. We first verify the computed fields with semi-analytical solutions for a simple layered model. Then we validate the responses of a more complex block model by comparing results obtained from each code. Finally we compare the responses of a real world model with results from the industry. On the one hand, these validations show that the open-source codes are able to compute comparable CSEM responses for challenging, large-scale models. On the other hand, they show many general and method-dependent problems that need to be faced for obtaining accurate results. Our comparison includes finite-element and finite-volume codes using structured rectilinear and octree meshes as well as unstructured tetrahedral meshes. Accurate responses can be obtained independently of the chosen method and the chosen mesh type. The runtime and memory requirements vary greatly based on the choice of iterative or direct solvers. However, we have found that much more time was spent on designing the mesh and setting up the simulations than running the actual computation. The challenging task is, irrespective of the chosen code, to appropriately discretize the model. We provide three models, each with their corresponding discretization and responses of four codes, which can be used for validation of new and existing codes. The collaboration of four code maintainers trying to achieve the same task brought in the end all four codes a significant step further. This includes improved meshing and interpolation capabilities, resulting in shorter runtimes for the same accuracy. We hope that these results may be useful for the CSEM community at large and that we can build over time a suite of benchmarks that will help to increase the confidence in existing and new 3D CSEM codes.
\end{summary}



\section{Introduction}

Controlled-source electromagnetic (CSEM) measurements are a frequently applied method in various areas of geophysical exploration, such as geothermal, groundwater, oil and gas, mining, civil engineering, or geohazards. Modelling these electromagnetic fields is therefore of great interest to design survey layouts, to understand the measured data, and for inversion purposes. Publications regarding three-dimensional (3D) modelling in electromagnetic (EM) methods started to appear as early as the 1970's and 1980's. These early publications were integral equation (IE) methods for simulating an anomaly embedded within a layered medium, mostly for loop-loop type transient EM measurements  \citep{GJI.74.Raiche, GEO.75.Hohmann, GJI.82.Das, GEO.86.Newman} and magnetotelluric (MT) measurements \citep{GEO.84.Wannamaker}. \cite{B.SEG.88.Ward} assemble many of these solutions in \emph{Electromagnetic Theory for Geophysical Applications}, which is widely viewed as an authoritative publication on electromagnetic geophysics.

In the 1990s, computers became sufficiently powerful that 3D modelling gained traction, and in 1995 the first International Symposium on Three-Dimensional Electromagnetics took place. This symposium resulted eventually in the book \emph{Three-Dimensional Electromagnetics} \citep{B.SEG.99.Oristaglio}, and another book by \cite{B.02.Wannamaker} with the exactly same title came out only three years later. Often cited publications from that time are \cite{RSC.94.Mackie} for 3D MT computation; \cite{RS.94.Druskin} for frequency- and time-domain modelling using a Yee grid and a global Krylov subspace approximation; and \cite{RS.96.Alumbaugh, GJI.97.Newman} for low-to-high frequency computation on massively parallel computers.

The continuous improvement of computing power and the CSEM boom in the early 2000's in the hydrocarbon industry led to a wealth of developed numerical solutions and according publications. The most commonly applied methods to solve Maxwell's equation are the IE method \citep{GJI.74.Raiche, RS.02.Hursan, GEO.06.Zhdanov, GP.10.Tehrani, CAG.16.Kruglyakov, MGS.17.Kruglyakov} and different variations of the differential equation (DE) method, such as finite differences (FD) \citep{GEO.93.Wang, RS.94.Druskin, RSC.94.Mackie, GEO.09.Streich, CAG.13.Sommer}, finite elements (FE) \citep{GEO.04.Commer, GJI.11.Schwarzbach,GEO.12.daSilva, GJI.13.Grayver, GJI.13.Puzyrev, SEG.16.Zhang}, and finite volumes (FV) \citep{EM.90.Madsen, SIAM.01.Haber, PIER.01.Clemens, GEO.14.Jahandari}. There are also many different types of discretization, where the most common ones are regular grids (Cartesian, rectilinear, curvilinear), mostly using a Yee grid \citep{IEEE.66.Yee} or a Lebedev grid \citep{CMMP.64.Lebedev}, but also unstructured tetrahedral grids \citep{SEG.16.Zhang, CAG.17.Cai}, hexagonal meshes \citep{CAG.14.Cai}, or octree meshes \citep{ECP.07.Haber}.

Very well written overviews about the different approaches to 3D EM modelling (and inversion) are given by \cite{SG.05.Avdeev} and \cite{SG.10.Borner}. But there was also a tremendous publications output with regards to 3D EM modelling in the last 10-15 years, at least partly driven by the ever increasing computing power. One reason why there are so many publications about this topic results from the variety of techniques to solve the systems of linear equations (SLE). They can be distinguished between direct solvers \citep{GEO.09.Streich, GP.14.Chung, GEO.14.Jaysaval, GEO.15.Grayver, SEG.15.Oh, GJI.18.Wang}, iterative solvers \citep{GP.06.Mulder, GJI.15.Jaysaval}, or a combination of both, so-called hybrid solvers \citep{GEO.18.Liu}. The solvers often use preconditioners such as the multigrid method \citep{SIAM.02.Aruliah, GJI.16.Jaysaval}.

Many of the advancements made in the EM modelling community over the past decades have required that authors develop new implementations from scratch. These codes often provided the research group or company with a competitive advantage for a time, and thus the source codes were normally kept internal. In some cases, executables have been made available for academic purposes upon request or to sponsoring consortium members. As the field continues to mature, advancements become more incremental, and particularly in an applied field, many advancements are driven by new use-cases and applications that were not considered by the original authors. In the aforementioned review on EM modelling and inversion Dmitry Avdeev concludes with the following statement: «\emph{The most important challenge that faces the EM community today is to convince software developers to put their 3-D EM forward and inverse solutions into the public domain, at least after some time. This would have a strong impact on the whole subject and the developers would benefit from feedback regarding the real needs of the end-users.}» Similarly, \cite{EXG.19.Oldenburg} argue that an open-source paradigm has the potential to accelerate multidisciplinary research by encouraging the development of modular, interoperable tools that are supported by a community of researchers.

Today it is becoming more common for researchers in many domains of science to release source-code with an open license that allows use and modification (e.g., see the Open Source Initiative approved licenses: \href{https://opensource.org/licenses}{opensource.org/licenses}). This is an important step for improving reproducibility of research, «\emph{to provide the means by which the reader can verify the validity of the results and make use of them in further research}» \citep{GEO.17.Broggini}. Going a step beyond releasing open-source software, many groups adopt an \emph{open model of development} where code is hosted and versioned in an online repository, all changes are public, and users can engage by submitting and track issues. Community oriented projects further engage by encouraging pull-requests, which are suggested changes to the code. Successful, well-maintained projects often have unit-testing and continuous integration that runs those tests with any changes to the code. Additionally, documentation that includes examples and tutorials is an important component for on-boarding new users and contributors. As a result, in many areas of the geosciences, we are seeing a shift away from a one-way distribution of (open-source) code towards building global communities around open projects. Other notable Python projects with an open model of development within the same realm as the codes under consideration are \texttt{pyGIMLi} \citep{CAG.17.Rucker}, \texttt{Fatiando} \citep{JOSS.18.Uieda}, \texttt{GemPy} \citep{GMD.19.DeLaVarga}, and \texttt{PyVista} \citep{JOSS.19.Sullivan}.

The landscape in the field of 3D CSEM modelling changed in the last five years quite dramatically. This paper introduces and compares four projects which make all source code openly available for use and adaptation. All presented codes are in the Python ecosystem and use either the FV method on structured grids or the FE method on unstructured tetrahedral meshes. Having different codes which use different methods and different meshes is ideal to address the topic of validation, and having all openly available facilitates reproducibility of our results. Analytical and semi-analytical solutions only exist for simple halfspace or layered-Earth models, which served mostly to verify new codes. The only objective possibility to ensure the accuracy of solutions beyond these simple models is by comparing results from different modellers. If different discretizations and implementations of Maxwell's equations yield the same result, it gives confidence in their accuracy. This is the principal motivation of our study, together with the necessity for more reproducible models and modelling results in the realm of 3D CSEM computations. \cite{GJI.13.Miensopust} presents a review of two workshops dealing with the validation of MT forward and inversion codes, but we are not aware of any comparable comparison study or benchmark suite for CSEM data.

We simulate EM fields for a layered background model with vertical transverse isotropy, with and without three resistive blocks, as well as for the complex marine, open-source MR3D model. These three models as well as the corresponding results from four different codes provide a benchmark for new (and existing) codes to be compared to and validated with. First we introduce the codes under consideration, and present afterwards the considered benchmark cases in detail together with the modelling results of our four codes in terms of accuracy and computational performance. Beyond that, we extensively discuss important points that control the performance and suitability of our FV and FE codes, including considerations about the mesh design and the choice of solvers. Such a comparison reveals avenues for further development for each of the codes. We conclude with a discussion and conclusions, and a motivation for the EM community at large to not only continue to extend the landscape of open-source codes but to also create a landscape of open-source benchmark models.

\section{Codes}

The four codes under consideration are, in alphabetical order, \custem \citep{GEO.19.Rochlitz}, \emg3d \citep{JOSS.19.Werthmuller}, \petgem \citep{CAG.18.CastilloReyes, GJI.19.CastilloReyes}, and \simpeg \citep{CAG.15.Cockett, CAG.17.Heagy}. All four codes have their user-facing routines written in Python, and all of them make heavy use of \texttt{NumPy} \citep{NAT.20.Harris} and \texttt{SciPy} \citep{NM.20.Virtanen}. The four of them follow the open model of development, meaning that they come with both an open-source license and an online-hosted version-control system with tracking possibilities (raising issues, filing pull requests). All developments comprise an extensive online documentation with many examples and have continuous integration to some degree. Newer package-management systems such as \texttt{conda}, \texttt{docker}, or \texttt{pip} simplify installation for all of these codes on any major operating system.

We briefly present our codes in this section. It is, however, beyond the scope of this article to go into every detail of the different modellers, and we refer to their documentations for more details. An overview comparison of the codes is given in Table~\ref{tbl:codecomparison}. All four have in common that they solve Maxwell's equation in its differential form under the quasi-static or diffusive approximation, hence neglecting displacement currents, which is a common approximation for the low frequency range usually applied in CSEM. For the numerical examples we show, all codes use the total-field formulation for the electric field. The machines on which the different codes were run are listed in Table~\ref{tbl:machines} together with the responsible operator. A few clarifying words on abbreviations and definitions: For the boundary conditions (BC) in the comparison table we use the abbreviations \emph{PEC} and \emph{PMC}, which stand for perfect electric conductor
and perfect magnetic conductor%
, respectively. PEC (PMC) implies that the tangential electric (magnetic) field vector components are zero at the boundary. Another abbreviation used in the tables is \emph{dof} for degree of freedom, which is equivalent to the size of the SLE we are solving. Finally, we use \emph{runtime} as the wall time or elapsed real-world time from start to end of solving the SLE; pre- and post-processing (e.g., mesh generation) is not measured. Note that the actual computation time or CPU time is therefore much higher than the reported runtime for the codes that run in parallel, with an upper limit of \#Procs$\times$Runtime. Memory refers to the maximum memory increase at any point of solving the SLE.

\begin{table*}
\begin{minipage}{.9\linewidth}
  \centering
  \caption{Comparison of the four codes under consideration. Note that \emg3d is a solver on its own, while the other codes implement third-package solvers such as \texttt{PETSc} \citep{Preprint.Abhyankar}, \texttt{MUMPS} \citep{SIAM.01.Amestoy}, or \texttt{PARDISO} \citep{FGCS.04.Schenk}.}
\label{tbl:codecomparison}
  \begin{tabularx}{\linewidth}{lXXXX}
  \toprule
  & \custem & \emg3d & \petgem & \simpeg  \\
  \midrule
  Website & \href{https://custem.rtfd.io}{custem.rtfd.io}
          & \href{https://emsig.github.io}{emsig.github.io}
          & \href{http://petgem.bsc.es}{petgem.bsc.es}
          & \href{https://docs.simpeg.xyz}{simpeg.xyz} \\
  License & GPL-3.0 & Apache-2.0 & BSD-3-Clause & MIT \\
  Installation & \texttt{conda}
               & \texttt{pip}; \texttt{conda}
               & \texttt{pip}
               & \texttt{pip}; \texttt{conda} \\
  Comp. Dom. & frequency \& time & frequency & frequency & frequency \& time \\
  Method & FE & FV & FE & FV \\
  Mesh & tetrahedral & rectilinear & tetrahedral & recti-/curvilinear, octree \\
  BC & PEC; PMC & PEC & PEC & PEC; PMC \\
  Solver & \texttt{MUMPS} & \emg3d & \texttt{PETSc}; \texttt{MUMPS} &
           \texttt{PARDISO}; \texttt{MUMPS} \\
  \bottomrule
\end{tabularx}
\end{minipage}
\end{table*}
\begin{table*}
\begin{minipage}{14cm}
  \centering
  \caption{List of hardware, software, and operator with which the different codes were run.}
\label{tbl:machines}
  \begin{tabularx}{\linewidth}{lXl}
  \toprule
  Code & Computer and Operating System & Operator \\
  \midrule
  \custem & \textbf{PowerEdge R940 (server)}                  & Raphael Rochlitz \\
          & - 144 Xeon Gold 6154 CPU @2.666 GHz               & \\
          & - $\approx3$ TB DDR4 RAM                          & \\
          & - Ubuntu 18.04                                    & \\[.5em]
  \emg3d  & \textbf{Dell Latitude (laptop)}                   & Dieter Werthmüller \\
          & - i7-6600U CPU@2.6 GHz x4                         & \\
          & - 15.5 GiB RAM                                    & \\
          & - Ubuntu 20.04                                    & \\[.5em]
  \petgem & \textbf{Marenostrum4 (supercomputer)}             & Octavio Castillo-Reyes \\
          & - 2 sockets Intel Xeon Platinum (Skylake generation) 8160 CPU & \\
          & \phantom{-} with 24 cores each @2.10GHz for a total of 48 cores per node & \\
          & - 386 GB DDR4 RAM per node                        & \\
          & - SuSE Linux Enterprise                           & \\[.5em]
  \simpeg & \textbf{GKE n2-custom (Google cloud)}             & Lindsey Heagy \\
          & - Intel Cascade Lake, 8 vCPUs                     & \\
          & - 420 GB RAM                                      & \\
          & - Ubuntu 16.04                                    & \\
  \bottomrule
\end{tabularx}
\end{minipage}
\end{table*}

\subsection{custEM}

The customisable electromagnetic modelling Python toolbox \custem was developed for simulating arbitrary complex 3D CSEM geometries with a focus on semi-airborne setups, but it supports also land-based, airborne, coastal, and marine environments. Multiple electric or magnetic field or potential finite-element approaches were implemented as total or secondary field formulations. The finite-element kernel, including higher order basis functions and parallelisation, relies on the \texttt{FEniCS} project \citep{B.SPR.12.Logg, B.SPR.16.Langtangen}. The resulting SLEs are solved with \texttt{MUMPS}, which is a very robust but memory consuming choice. Primary field solutions are supplied by the \texttt{COMET} package \citep{GEO.20.Skibbe}.

The toolbox considers generally anisotropic petrophysical properties. Even though changes of the conductivity are mainly of interest for CSEM modelling, the electric permittivity and magnetic permeability can be taken into account using the preferred electric field approach on Nédélec elements. Recently, induced polarisation parameters in frequency-domain computations and three methods for simulating time-domain responses were added to \custem. The provided meshing tools are based on \texttt{TetGen} \citep{TOM.15.Si} and functionalities of \texttt{pyGIMLi} facilitate the generation of tetrahedral meshes including layered-earth geometries with topography or bathymetry and anomalous bodies which are allowed to be connected or to reach the surface.

\subsection{emg3d}

The 3D CSEM code \emg3d is a multigrid solver \citep{CMMP.64.Fedorenko} for electromagnetic diffusion following \cite{GP.06.Mulder}, including tri-axial electrical anisotropy. The matrix-free solver can be used as main solver or as preconditioner for one of the Krylov subspace methods implemented in SciPy. The governing equations are discretized on a staggered grid by the finite-integration technique \citep{AEU.77.Weiland}, which is a finite-volume generalisation of a Yee grid. The code is written completely in Python using the \texttt{NumPy} and \texttt{SciPy} stacks, where the most time- and memory-consuming parts are sped up through jitted (just-in-time compiled) functions using \texttt{Numba} \citep{LLVM.15.Lam}. The strength of \emg3d is the matrix-free multigrid implementation, which is characterised by almost linear scaling both in terms of runtime and memory usage, and it is therefore a solver that uses comparably very little memory.

As such the code is primarily a solver, which solves Maxwell's equations for a single frequency and a single, electric source. Recent developments added functionalities such that \emg3d can be used directly as a more general EM modeller too. In addition to functionality for modelling arbitrarily rotated electric sources and receivers, it also can simulate magnetic sources and receivers as well as model time-domain responses. It further has routines for obtaining the gradient of the misfit function which can be used by inversion algorithms. For the underlying discretization the Python package \discretize is used, which is part of the larger \simpeg ecosystem.

\subsection{PETGEM}

\petgem is a parallel code for frequency-domain 3D CSEM data for marine and land surveys. The high-order edge FE method (HEFEM) is used to discretize the governing equations in its diffusive form. This technique provides a suitable mechanism to obtain stable numerical solutions and a good trade-off between the number of dof and computational effort. The current implementation supports up to sixth-order tetrahedral vector basis functions. Moreover, because the HEFEM belongs to the FE family, the unstructured meshes can be used efficiently on complex geometries (e.g., models with topography and bathymetry).

\petgem permits to locate the source and receivers anywhere in the computational domain (e.g., sediments, seafloor, sea, ground), allowing to analyse the physical environment of the electric responses and how parameters impact them (e.g., frequency, conductivity, dependence on mesh setup, basis order, solver type). Furthermore, \petgem implements a semi-adaptive mesh strategy ($hp$ mesh refinement) based on physical parameters and on polynomial order to satisfy quality criteria chosen by the user. Nonetheless, only Horizontal Electric Dipole (HED) has been implemented.

For the parallel forward modelling computations, a highly scalable MPI (message passing interface) domain decomposition allows reducing runtimes and the solution of large-scale modelling cases. This strategy is capable of exploiting the parallelism offered by both modest multi-core computers and cutting-edge clusters such as High-Performance Computing (HPC) architectures.

\subsection{SimPEG}

\simpeg is a modular toolbox for simulations and gradient based inversions in geophysics. Current functionality includes modelling and inversion capabilities for gravity, magnetics, direct current resistivity, induced polarisation, EM (time and frequency domain, controlled and natural sources), and fluid flow. It is a community driven project that aims to support researchers and practitioners by providing a flexible, extensible framework, and common interface to a variety of geophysical methods.

Meshes and finite volume differential operators are implemented in the \discretize package, which currently includes rectilinear, octree, cylindrical, and logically rectangular meshes. Each mesh type inherits from a common structure and uses the same naming conventions for methods and properties. This allows us to decouple the implementation of a discretized partial differential equation from the details of the mesh geometry and therefore we can write a single implementation of discretized partial differential equation (or set of partial differential equations) in \simpeg that will support all mesh types implemented in \discretize.

The electromagnetic implementations use a staggered grid finite volume approach where physical properties are discretized at cell centres, fields on cell edges and fluxes on cell faces. There are two different discretization strategies implemented for Maxwell's equations: (I) the EB-formulation, which discretizes the electric field ($\vec{e}$) and the magnetic flux density ($\vec{b}$) and (II) the HJ-formulation, which discretizes the magnetic field ($\vec{h}$) and the current density ($\vec{j}$). Having multiple implementations allows for testing that compares results from each approach, as well as the representation of both electrical and magnetic sources on cylindrically symmetric meshes. \simpeg supports variable magnetic permeability and full-tensor anisotropy for the physical properties. \simpeg interfaces to various solvers including \texttt{PARDISO} and \texttt{MUMPS}, and also to some implemented in \texttt{SciPy}.

\section{Numerical Validation}

We computed the responses for three different models to validate that the four 3D CSEM codes yield the same electromagnetic responses and compare different solver types (FE, FV) and different mesh types (unstructured tetrahedra, rectilinear, octree) in different scenarios. The first model is an anisotropic, layered model, which can be verified with semi-analytical solutions. The layered model also serves as background model for the second validation, where we add three resistive blocks into the subsurface. The final comparison is based on the realistic, anisotropic marine model MR3D. It is a large-scale resistivity model that is openly released, together with computed CSEM responses.

\subsection{Layered Model}

The layered (1D) model consists of an upper halfspace of air ($\rho_\text{air}=\num{e8}\,\ohmm$), a 600\,m deep water layer ($\rho_\text{sea}=0.3\,\ohmm$), followed by a 250\,m thick, isotropic layer of 1\,\ohmm, a 2.3\,km thick, anisotropic (vertical transverse isotropy, VTI) layer of $\rho_\text{h}=2\,\ohmm$ and $\rho_\text{v}=4\,\ohmm$, and finally a resistive, isotropic basement consisting of a lower halfspace of $1000\,\ohmm$. The survey uses a 200\,m long, 800\,A source  with frequency $f=1\,$Hz at a single position, and three receiver lines of 101 receivers each. The centre of the $x$-oriented source is at $x=y=0\,$m, 50\,m above the seafloor. The $x$-directed receivers are placed on the seafloor every 200\,m from $x=-10\,$km to $x=+10\,$km in three lines with $y=-3; 0; +3\,$km.

\begin{figure*}
  \centering
  \includegraphics[width=.9\fwidth]{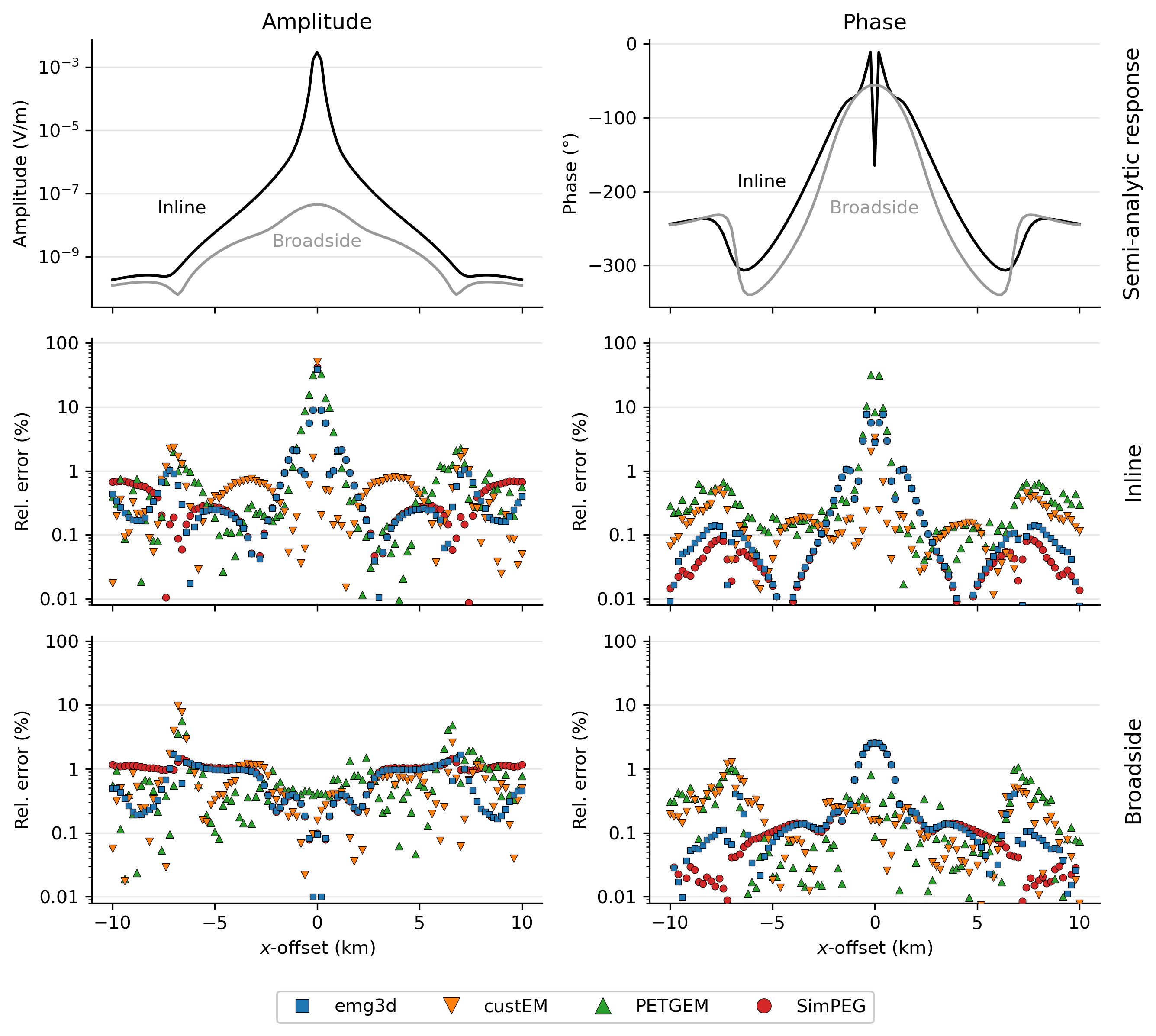}
  \caption{Results of the layered model: The semi-analytical inline and broadside responses are shown in the top row. The relative percentage error of the four 3D modellers are shown in the second and the third row for the inline and broadside responses, respectively.}
  \label{fig:results-layered}
\end{figure*}

A layered model fails to show the strength of 3D modellers in general and FE codes in particular, and in reality one would not choose a 3D code to compute responses for such a problem. However, it is one of the few examples that can be verified with semi-analytical results, and it is therefore an important first test. The results are shown in Figure~\ref{fig:results-layered}, in the left column the amplitudes and in the right column the phases. The top row shows the actual, semi-analytically computed responses, for which we used \empymod\citep{GEO.17.Werthmuller}. The second and the third row show the relative percentage error of the inline and the broadside receivers, respectively; note that for the 1D case the x-directed electric fields for the two broadside lines for $y=\pm3\,$km are identical.
The misfit is generally in the order of 1\,\% or less, the biggest misfits are on the one hand close to the source and on the other hand where there are rapid phase changes associated to amplitudes having a dip towards zero. The increased errors towards the location of the 200\,m long source are related to the gridding, and finer discretizations (resulting in longer runtime and higher memory consumption) would reduce these errors.

Note that \emg3d and \simpeg used the same rectilinear mesh in this example. The use of an octree mesh in \simpeg could greatly reduce the size of the mesh and resultant computation time, but at the cost of accuracy, which is reduced when averaging over layer interfaces to coarsen cells and because of the factor-of-two expansion rate of coarsened cells in an octree mesh. As the aim of this study is to focus on accuracy of the simulations and compare discretization approaches, we chose to use the same mesh as in the \emg3d simulation. Although we only show the x-directed field in this comparison, the y-directed field yields a similar result with respect to the misfit. A different story in this scenario is the z-directed field, which is not continues across the boundary on the seafloor. Having receivers close to such a non-continues boundary requires advanced interpolation routines \citep[e.g., ][]{GJI.15.Shantsev, GP.11.Wirianto}. The FE codes handle this problem automatically with the used basis functions, but the used FV codes do currently not have such routines implemented. The grid for the FV codes in this first example with a focus on precision is created in such a way that the receivers are located on the corresponding nodes, so no interpolation is required to obtain the responses.

%

\subsection{Block Model}

\begin{figure}
  \centering
  \includegraphics[width=\cwidth]{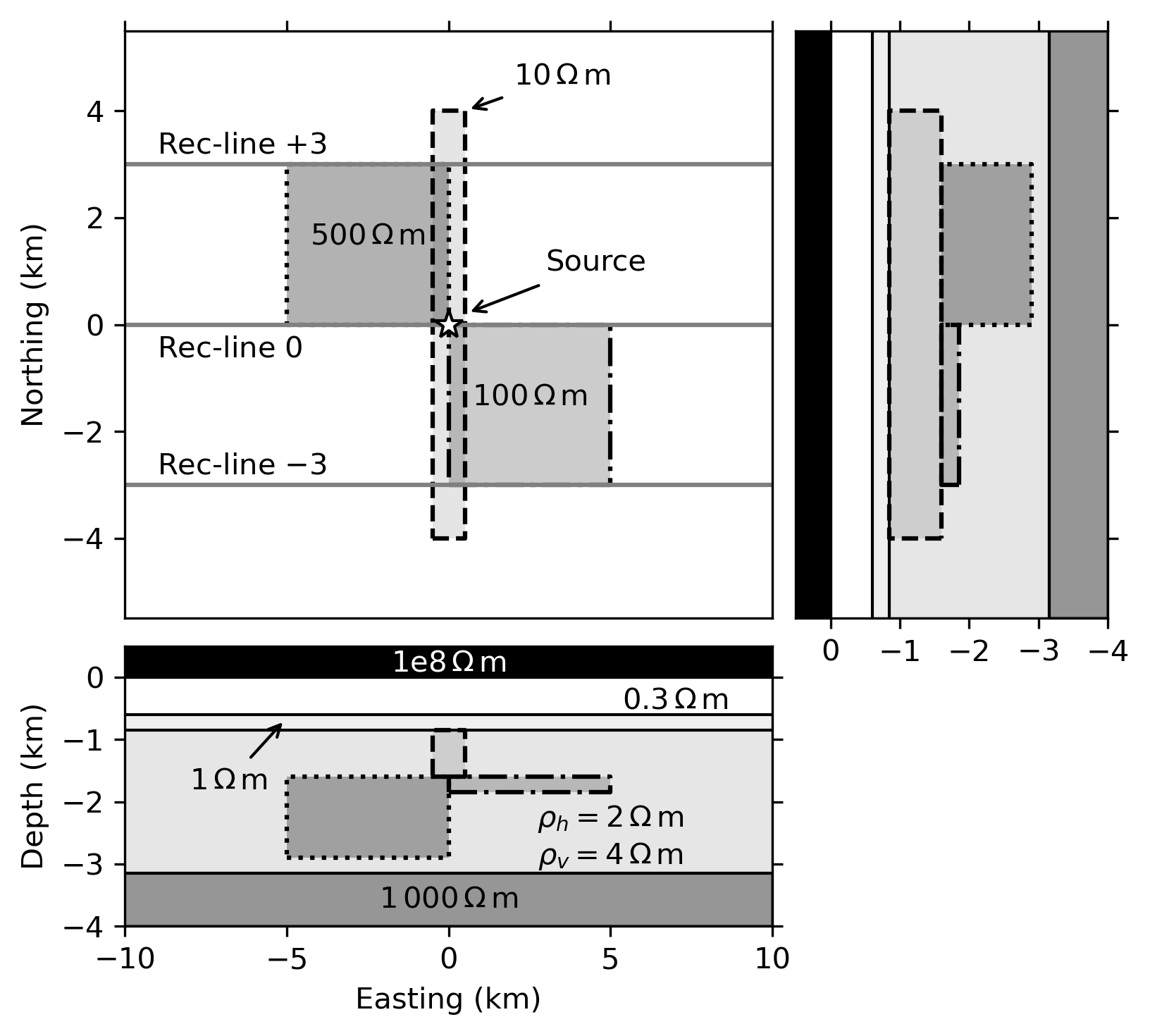}
  \caption{Sketch of the block model, consisting of the layered background model with three resistive blocks, embedded in the thick background layer which has VTI with $\lambda=\sqrt{\rho_\textrm{v}/\rho_\textrm{h}}=\sqrt{2}$.}
  \label{fig:model-block}
\end{figure}

The block model is a CSEM adaption of the MT \emph{Dublin Test Model~1} from the first EM modelling workshop described by \cite{GJI.13.Miensopust}. We use the same layout of the blocks but adjust the dimensions and resistivities to a typical marine CSEM problem, as shown in Figure~\ref{fig:model-block}. Additionally, we add our \emph{Layered Model} as a VTI background. The three resistive blocks have resistivities of $10\,\ohmm$ (shallow beam perpendicular to survey lines), $100\,\ohmm$ (thin plate, South-East), and $500\,\ohmm$ (cube, North-West).

There are no semi-analytical solutions for such a model to verify our results, and we can only validate them by comparing different results. For this we use the normalised root-mean square difference (NRMSD) between two responses $R_1$ and $R_2$, given by
\begin{equation}
  \mr{NRMSD~(\%)} = 100\,\frac{|R_1 - R_2|}{(|R_1| + |R_2|)/2} \ ,
  \label{eq:nrmsd}
\end{equation}
to which we refer as simply the \emph{normalised difference} throughout the paper.

\begin{figure*}
  \centering
  \includegraphics[width=.9\fwidth]{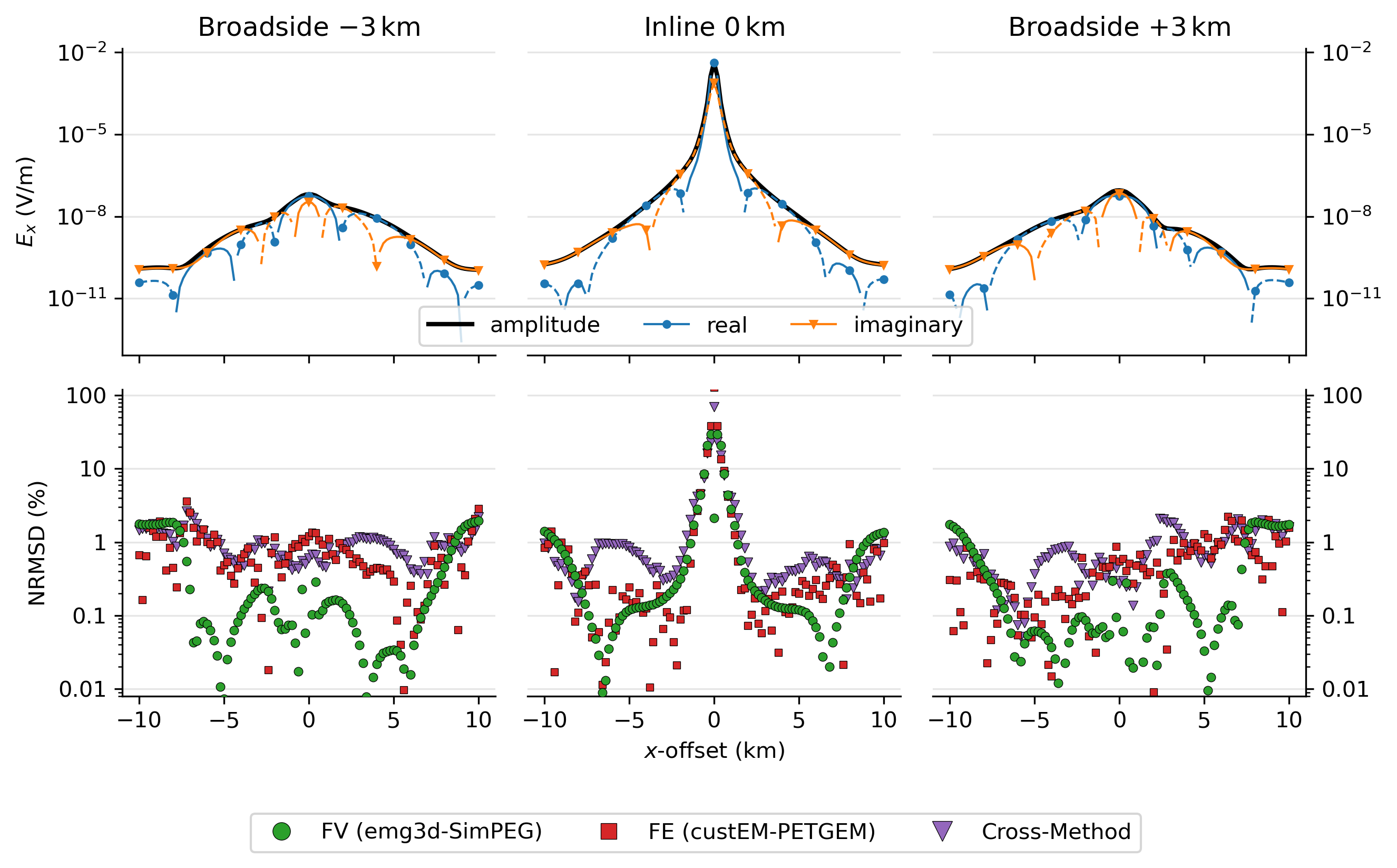}
  \caption{Results of the block model comparison: The responses of \emg3d, as an example, are shown in the top row, and the normalised differences (\%) between the amplitudes of the different codes are shown in the bottom row.}
  \label{fig:results-block}
\end{figure*}

The results for the three receiver lines $y=-3;0;+3\,$km are shown in the left, middle, and right columns of Figure~\ref{fig:results-block}, respectively. The top row shows the result of \emg3d, as an example, and the bottom row shows the normalised differences between the absolute responses of the codes. Note that for visual reasons we only show the normalised differences between the two FV codes, the two FE codes, and between \emg3d and \custem as an example of a cross-FV-FE comparison; the three other cross-method comparisons look quantitatively similar.

A few interesting points stand out:
\begin{enumerate}
  \item The misfit levels between the different codes are overall comparable; the normalised difference is generally below a few percent, increasing towards larger offsets.
  \item The normalised difference between the FV codes \emg3d and \simpeg is comparably low except at the boundary. The generally low normalised difference is because these codes use the same rectilinear mesh for modelling. The increasing normalised difference towards the boundary is because \emg3d uses the PEC boundary condition, while \simpeg uses the PMC boundary condition in this example. This difference from the boundary condition can also be seen in the relative error of the layered model in Figure~\ref{fig:results-layered}.
  \item The FE codes \custem and \petgem show similar boundary effects, but are influenced by the anomaly response (higher misfits over the anomalous blocks).
  \item The agreement within the same differential equation method (FE or FV) is mostly better than across methods.
\end{enumerate}

The corresponding required runtime and memory are listed in Table~\ref{tbl:comp-block}.
\begin{table*}
\begin{minipage}{10cm}
  \centering
  \caption{Comparison of number of processes, runtime, and memory, as well as the degree of freedom of the discretization used by the different codes for the block model.}
\label{tbl:comp-block}
  \begin{tabular}{lrS[table-format=6.0]S[table-format=4.1]S[table-format=8.0]}
  \toprule
  Code & \#Procs & {Runtime (s)} & {Memory (GiB)}   & {\#dof} \\
  \midrule
  \custem & 48 &   312 & 281.8 & 6014440 \\ 
  \emg3d  &  1 &   213 &   0.5 & 6004144 \\
  \petgem & 24 &   238 & 152.8 & 2455868 \\
  \simpeg &  4 & 20000 & 387.9 & 6004144 \\ 
  \bottomrule
\end{tabular}
\end{minipage}
\end{table*}

\subsection{Marlim R3D}

The Marlim oil field is a giant reservoir in a turbidite sandstone horizon in the north-eastern part of the Campos Basin, offshore Brazil, which was discovered in 1985. \cite{BJG.17.Carvalho} created from seismic data and well log data a realistic, three-dimensional resistivity model with vertical transverse isotropy (VTI), called MR3D, which they released under the open creative common license \emph{CC BY 4.0}. \cite{GEO.19.Correa} computed CSEM data for MR3D for six frequencies from 0.125\,Hz to 1.25\,Hz, and released them under the same CC license. To compute the data they used a code from the industry \citep[][ \emph{SBLwiz} software from \emph{EMGS}]{GEO.07.Maao}. It is therefore, on the one hand, an ideal case to validate our open-source codes against, as it is a complex, realistic model and the data were computed by an industry-proofed code. On the other hand it is impossible to reproduce it exactly, as it is a closed-source code and we cannot know exactly what was done internally. Additionally, that code is a time-domain code obtaining the frequency-domain responses through a transform, and it includes the air layer via a non-local boundary condition at the water-air interface \citep{GEO.10.Mittet}. We compute all the results in this study in the frequency domain, where we model the entire domain including the air layer by putting the boundaries of the computational domain far away.

\begin{figure*}
  \centering
  \includegraphics[width=.9\fwidth]{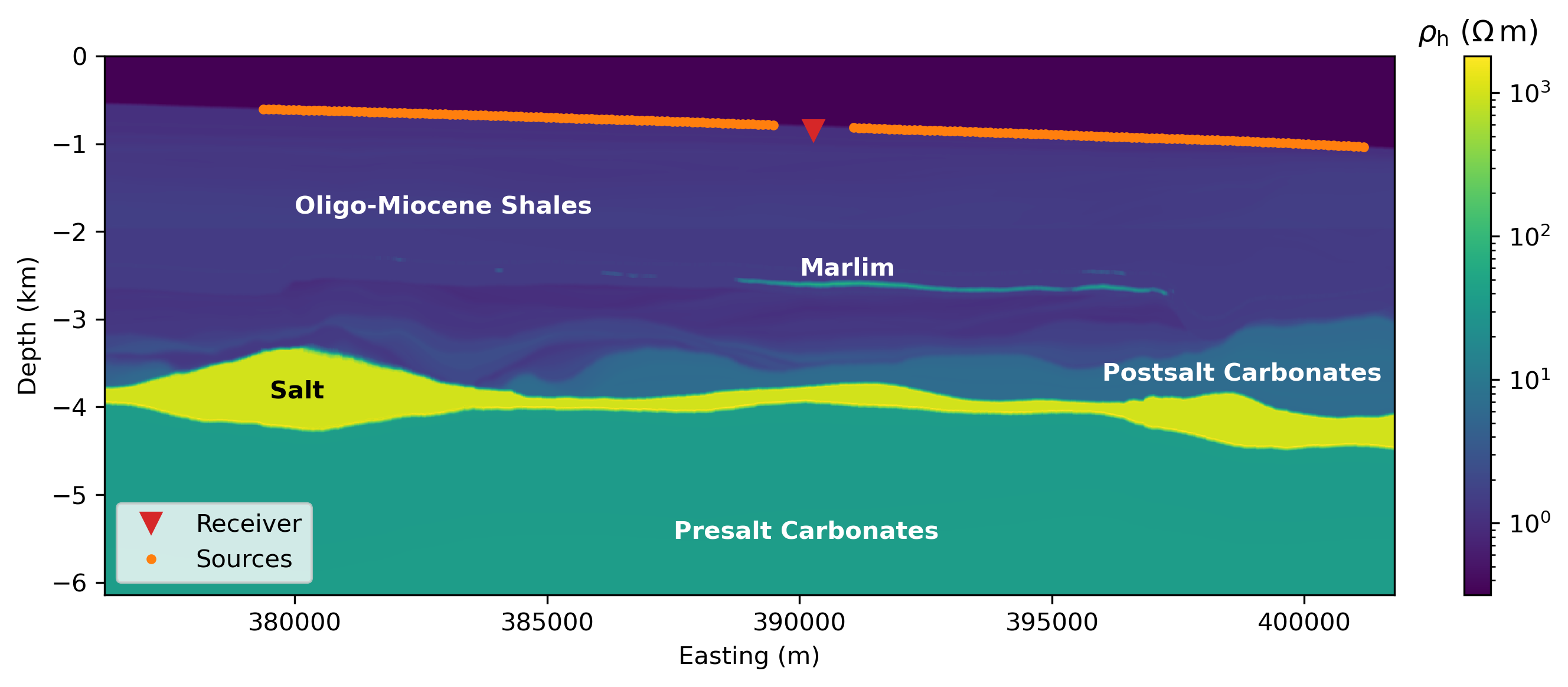}
  \caption{MR3D horizontal resistivity model, $x$-$z$-slice through the receiver $y$-position and with major formations annotated. Air (not shown in the model), seawater, and the salt layer are electrically isotropic, everything else has VTI with $\lambda=\sqrt{2}$. The receiver is located on the seafloor, and the sources fly 50\,m above the seafloor.}
  \label{fig:model-marlim}
\end{figure*}

The full MR3D model consists of $1022\times371\times1229$ cells, totalling to almost 466 million cells, where each cell has dimensions of $25\times75\times5$\,m. The model was upscaled for the computation to $515\times563\times310$ cells, totalling to almost 90 million cells, where each cell has dimensions of $100\times100\times20$\,m. Both the full model and the upscaled computational model are released as open source. The published data set consists of a regular grid of receivers of 20 in eastern direction by 25 in northern direction, 500 receivers in total, with 1\,km spacing located on the irregular seafloor. 45 source-towlines were located on the same grid above each receiver line, 50\,m above the seafloor, with shots every 100\,m. The computed responses for one of the receivers, \texttt{04Rx251a}, are shown in the original publication for the East-West inline-source \texttt{04Tx013a} and the East-West broadside-source \texttt{04Tx014a} (broadside offset of 1\,km). We reproduce the responses for this receiver and corresponding source-lines in our comparison. The $x$-$z$ cross-section of the horizontal resistivity model at the receiver position is shown in Figure~\ref{fig:model-marlim}, together with the receiver and sources positions. All layer have a VTI with $\lambda=\sqrt{2}$, except for the air layer, the seawater, and the salt layer, which are all isotropic.

\begin{figure*}
  \centering
  \includegraphics[width=.9\fwidth]{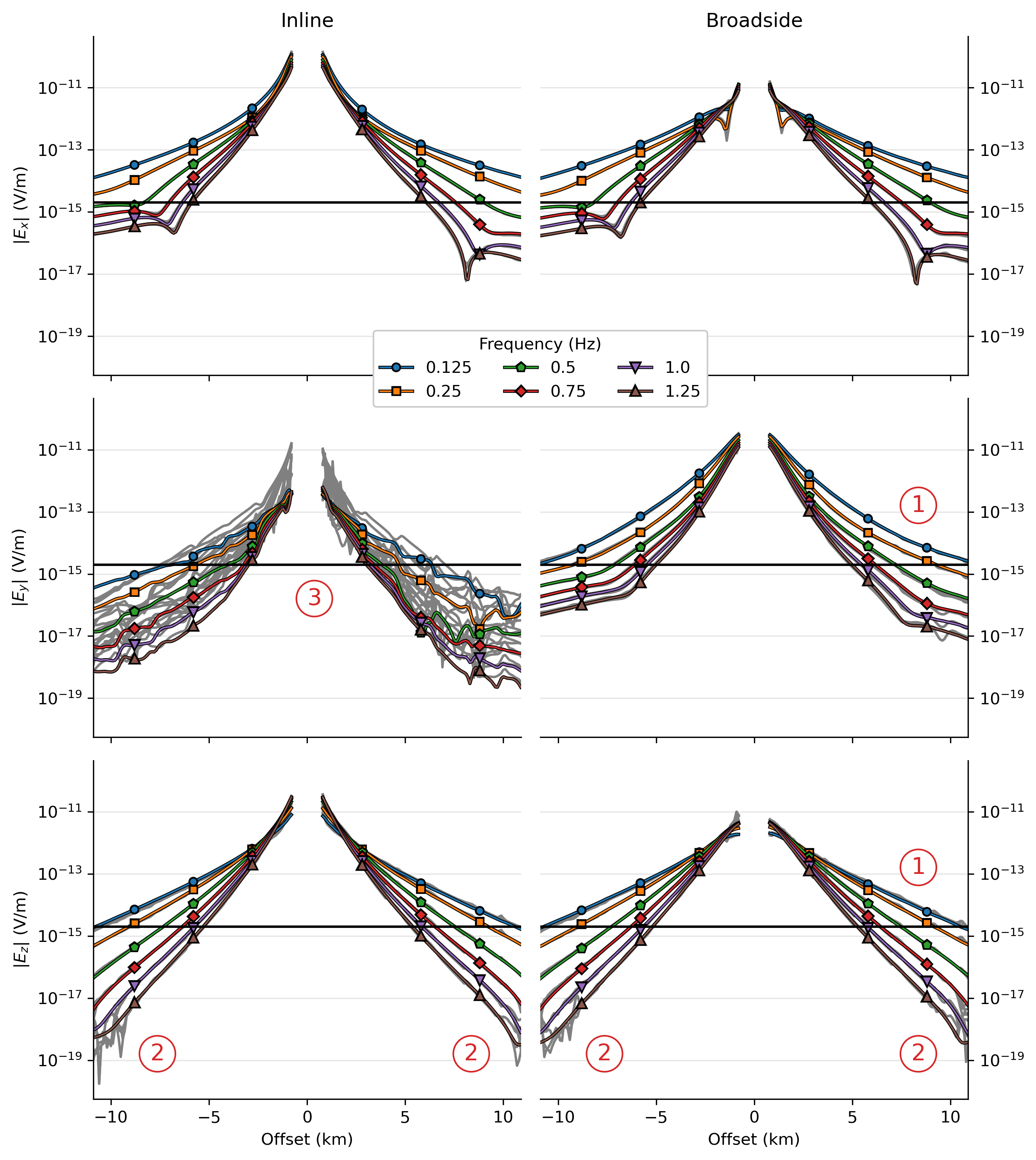}
  \caption{MR3D comparison between our codes (grey lines) and the published data (coloured lines). The grey lines under the coloured lines are not visible in most areas, meaning that they are very similar. However, there are three notable zones, (1) to (3), which are explained further in the text.}
  \label{fig:results-marlim-responses}
\end{figure*}

\begin{figure*}
  \centering
  \includegraphics[width=.9\fwidth]{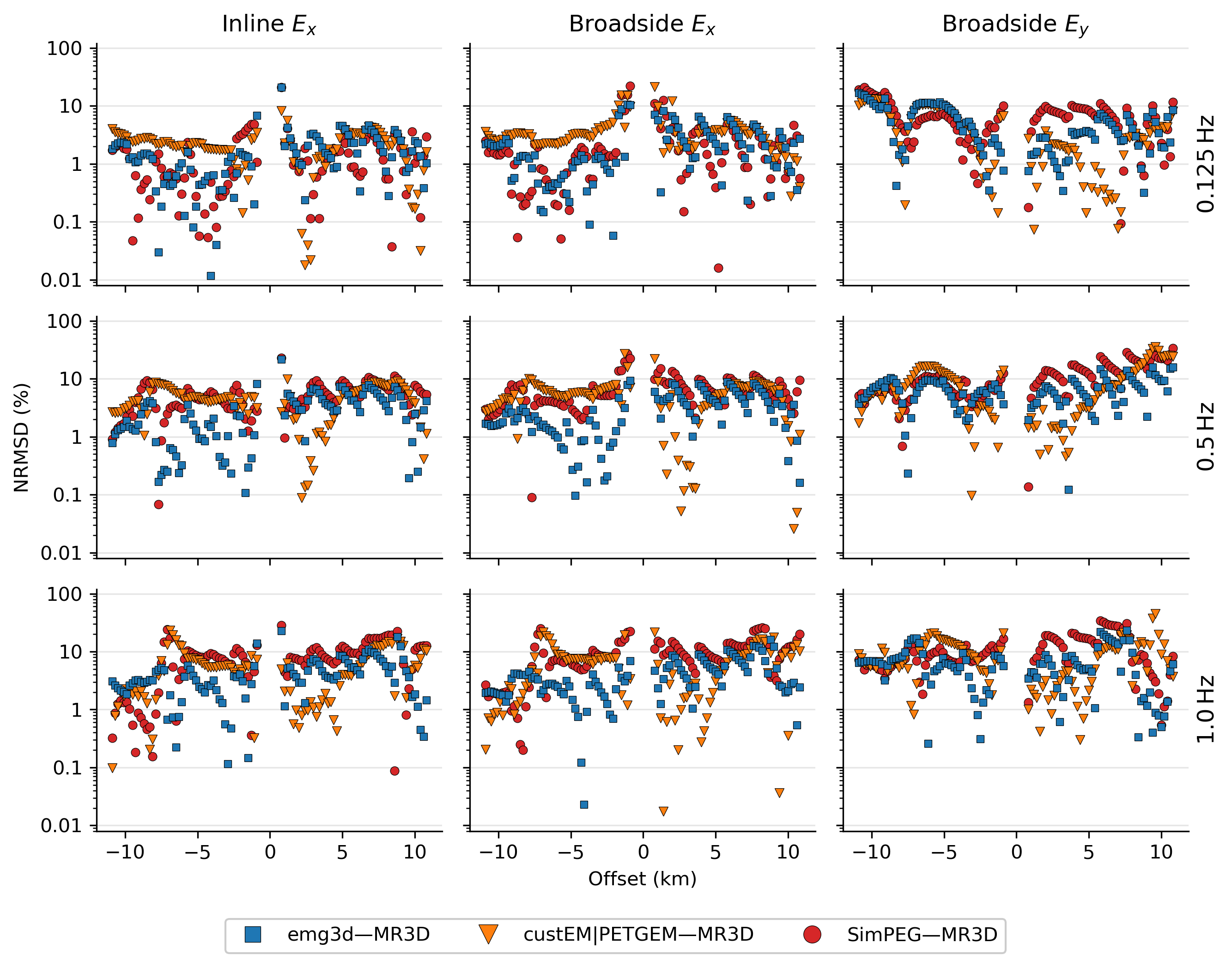}
  \caption{Normalised differences for all four codes in comparison with the published data, for the inline $E_x$ field and the broadside $E_x$ and $E_y$ fields in the left, middle, and right column, respectively. Shown are the three frequencies 0.125\,Hz, 0.5\,Hz, and 1.0\,Hz in the top, middle, and bottom row, respectively. The results of the two FE codes are almost identical, see Figure~\ref{fig:results-marlim_2ours}, and are therefore combined here.}
  \label{fig:results-marlim_2published}
\end{figure*}

The responses for all six frequencies and all three electric components are shown in Figure~\ref{fig:results-marlim-responses}, both inline and broadside (we only show the amplitude, as the conclusions for the phase are very similar). The published responses are shown in colour with markers for the frequency, and the responses from our codes are shown beneath in grey colours in this overview plot.
Two CSEM data-sets were published with MR3D, one containing clean responses, and one responses where realistic noise was added. For the comparison we use the clean data, but we indicate the chosen noise level by the horizontal line at \num{2e-15}\,V/m. It can be seen that the grey lines are not visible for most parts, which means that the data agree well. However, there are a few notable points where this is not the case, and they are marked with numbers in the figure.
\begin{enumerate}
  \item[(1)] There are some noticeable differences for positive offsets of the broadside $E_y$ and $E_z$ components, which are likely related to the bathymetry. The reason why we cannot reproduce the published results exactly is that the code is not open-source, and we do not know in detail what it does internally. See the discussion around Figure~\ref{fig:results-marlim_2published}.
  \item[(2)] The three highest frequencies of the $E_z$ fields become noisy at large offsets for all of our codes (well below any real-world noise level).
  \item[(3)] The responses for the $E_y$ component of the inline acquisition line do not agree at all. In the 1D case, the inline $E_y$ component would be zero, and the only response we can measure here comes from 3D effects. These responses are very low, roughly two orders of magnitude lower than the $E_x$ component. Not the published nor our responses are stable, and any of the responses is as bad as the other. Tiniest differences in meshing or different interpolation algorithms will have a huge effect here.
\end{enumerate}

It is important to note that there is no \emph{true solution}; but the closer that the previously published results and the results obtained from the four codes run here agree, the more confident we can be of the outcomes. This emphasises the importance of comparing 3D results and one of our main objectives, as there is no other way to check the results of 3D codes for complex models. Figure~\ref{fig:results-marlim_2published} shows the normalised differences between the published results and our codes for the three strongest components, the inline $E_x$ and the broadside $E_x$ and $E_y$ components, for three frequencies.
A few conclusions can be drawn from this figure: In general the normalised difference is roughly 10\,\% or less. The structured meshes have in some cases a smaller normalised difference, e.g., for negative offsets in the inline and broadside $E_x$ responses, and in other cases the unstructured tetrahedral meshes, e.g., for positive offsets for all broadside $E_y$ responses, and often they have a comparable NRMSD. Some normalised differences have an interesting step-pattern, which is more pronounced for the structured meshes than for the unstructured meshes; this is related to the annotated point (1) in Figure~\ref{fig:model-marlim}. The origin of this zigzag pattern is most likely the bathymetry, as the code of the published data uses the bathymetry information in addition to the resistivity cube. This explains why the codes using the tetrahedra mesh, which use the bathymetry as well, do not show this feature. None of our codes has currently special interpolation routines implemented to take particular care of a source injection close to a dipping interface, such as the bathymetry, with discontinuities in the normal electrical field and the normal derivatives of both electric and magnetic tangential fields. This effect alone can potentially contribute several percents of NRMSD, as shown by \mbox{\cite{GJI.15.Shantsev}}.

\begin{figure*}
  \centering
  \includegraphics[width=.9\fwidth]{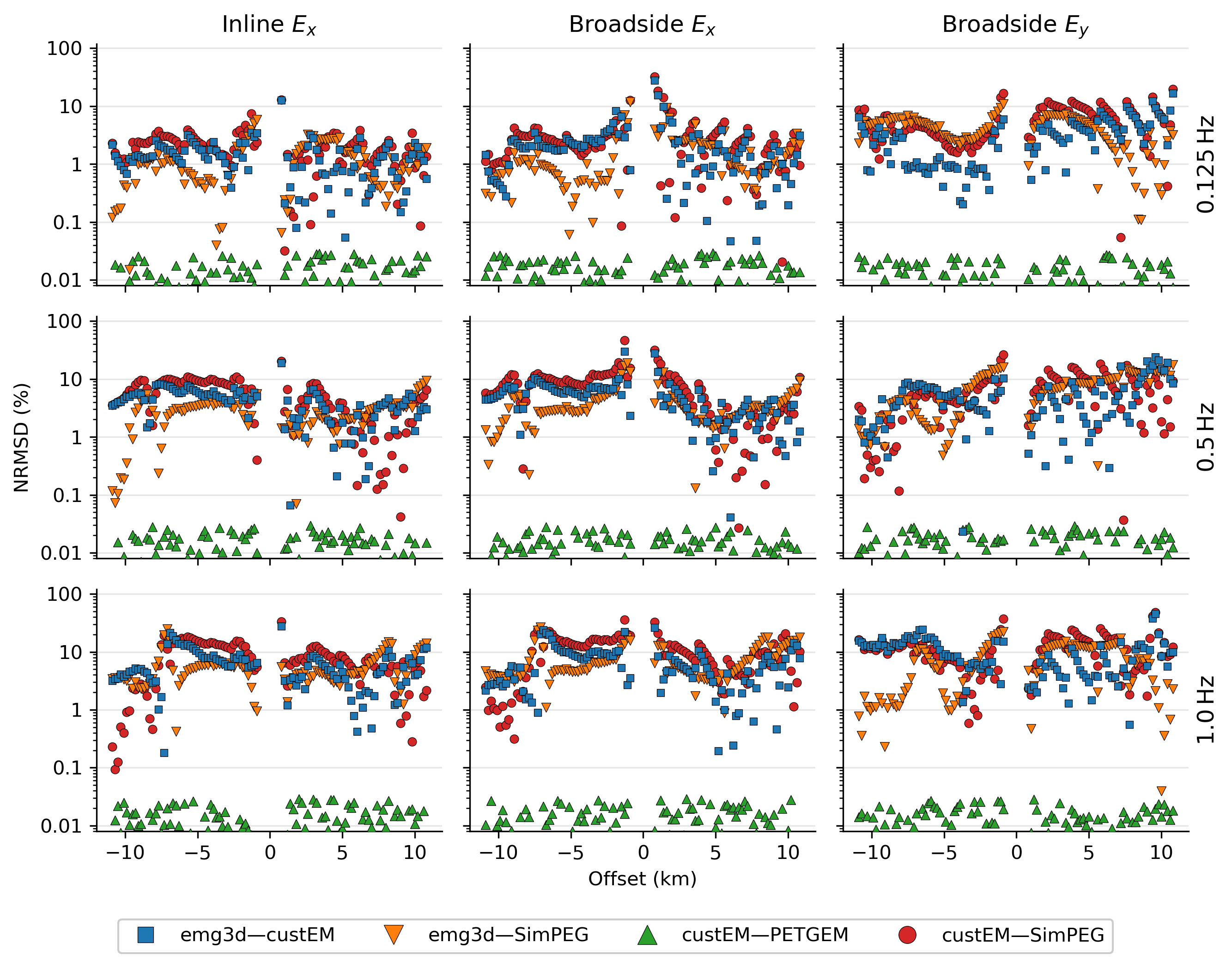}
  \caption{Normalised differences, just as in Figure~\ref{fig:results-marlim_2published}, but comparing some of the four codes with each other. \custem and \petgem produce very similar results, which is due to the fact that they use the same mesh. This also shows that the biggest impact in 3D forward modelling comes from the discretization.}
  \label{fig:results-marlim_2ours}
\end{figure*}

The importance of the meshing to the result can be seen in Figure~\ref{fig:results-marlim_2ours}, where we compute the normalised differences between our codes.
The most obvious result in that figure is that \custem and \petgem produce results that are almost identical, their normalised difference is generally below 0.03\,\%. This is why they are shown as one case in Figure~\ref{fig:results-marlim_2published}. The reason is simple: \custem and \petgem use the exactly same mesh for the computation of this example, the only difference is therefore the solver. This exemplifies that the biggest difference between different codes is the discretization; the solver is important when it comes to runtime and memory, but not for the accuracy of the result. As these two are so similar we compare \emg3d and \simpeg only to \custem in Figure~\ref{fig:results-marlim_2ours}, as the comparisons to \petgem would look the same. But in general the normalised differences between our codes look similar as in the comparison with the published results, and the principal source of the difference must be in the different discretizations.

\begin{figure*}
  \centering
  \includegraphics[width=.9\fwidth]{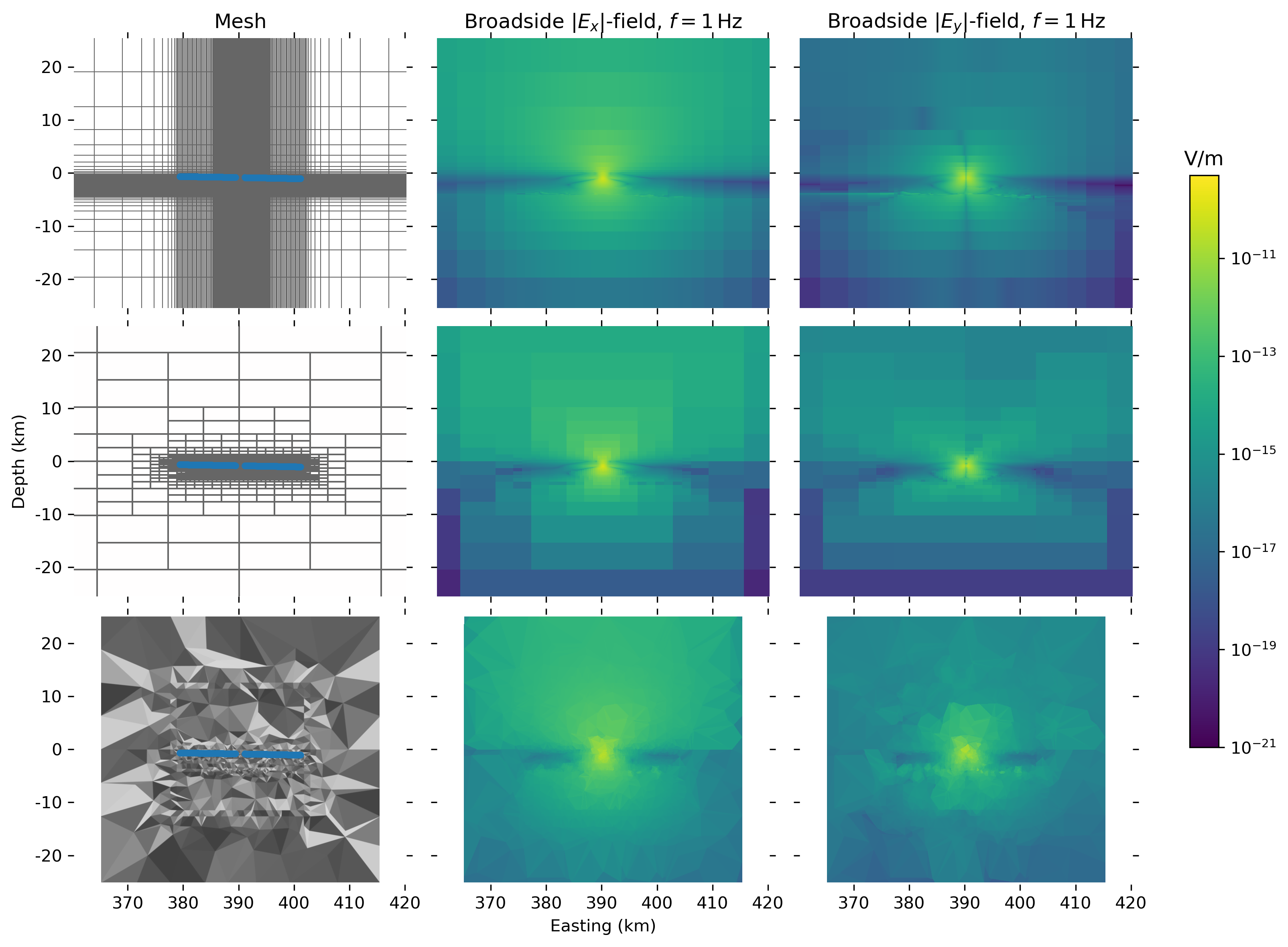}
  \caption{Meshes and broadside $E_x$ and $E_y$ fields in the left, middle, and right column, respectively, for the rectilinear, octree, and tetrahedral meshes in the top, middle, and bottom row, respectively. The shown dimensions display the entire computational domain for the tetrahedra mesh, whereas the entire rectilinear and octree meshes are bigger.}
  \label{fig:results-marlim_big}
\end{figure*}

The actual differences in meshing is shown in the next two figures. Figure~\ref{fig:results-marlim_big} shows on the top row the rectilinear mesh of \emg3d, on the middle row the octree mesh of \simpeg, and on the bottom row the tetrahedral mesh of \custem and \petgem. The actual mesh and the broadside $|E_x|$ and $|E_y|$ fields are shown in the left, middle, and right columns, respectively. The extent of the plot includes the entire computational domain for the tetrahedral mesh (in x and z direction); the rectilinear and octree mesh extents are larger than what is shown (rectilinear: $x=327$ to 454\,km, $z=-57$ to 66\,km; octree: $x=288$ to 492\,km, $z=-123$ to 41\,km).
This figure is interesting for various reasons, as the differences between the three meshes are clearly visible. Particularly the advantage of octree and tetrahedra to create large cells outside of the region of interest to push away the boundary can be seen, whereas a rectilinear grid, even though stretched, has to compute many cells far away of the zone of interest. The different boundary condition also play a part, where the top row used PEC, and the other two rows PMC. Also visible is that there is more energy close to the boundary in the tetrahedral mesh, due to the smaller computational domain with corresponding reflections. However, this seems not to have influenced the responses at the receiver locations, as the misfit between the FE and FV codes does not increase particularly for receivers with larger offsets.

\begin{figure*}
  \centering
  \includegraphics[width=.9\fwidth]{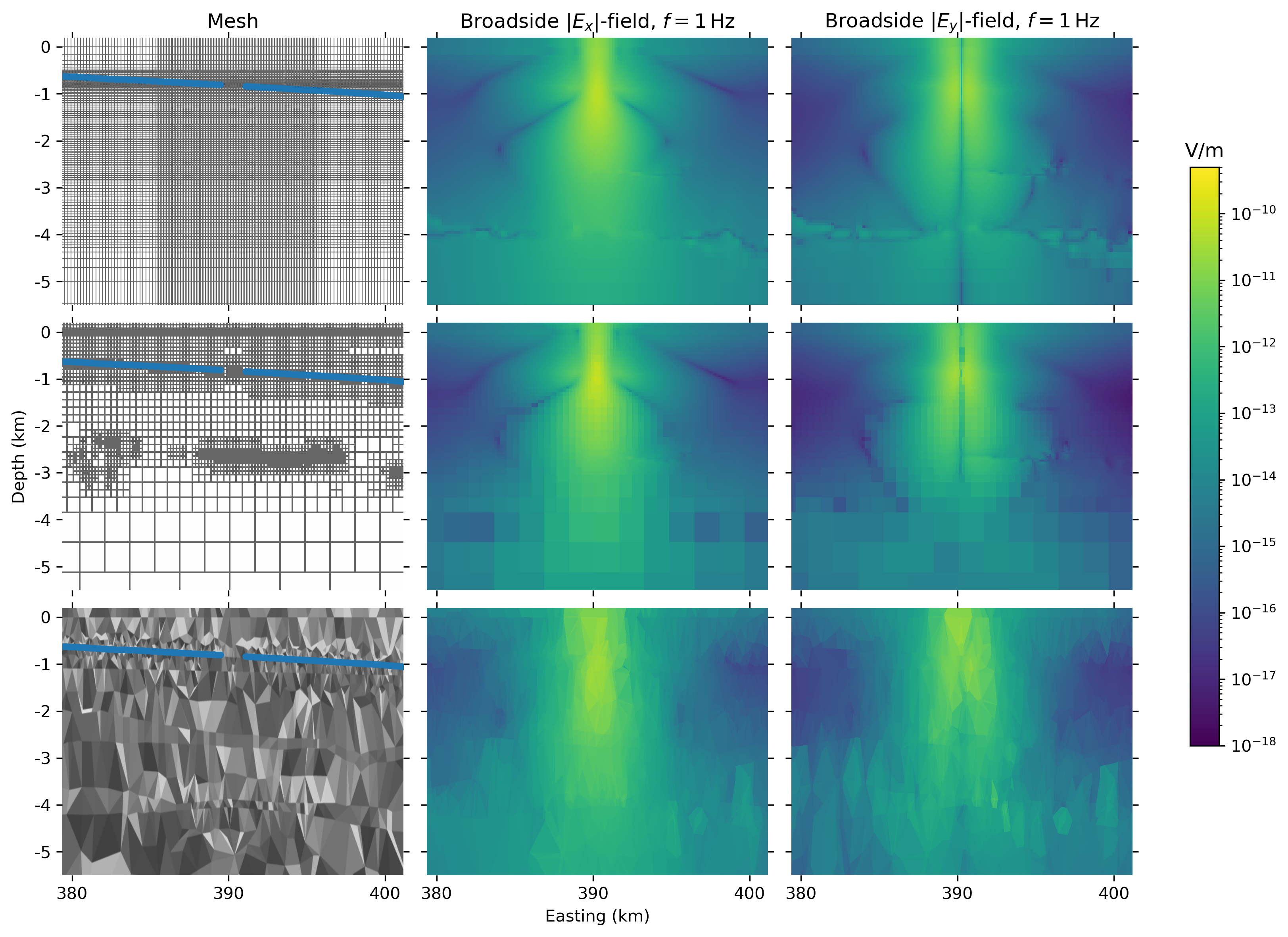}
  \caption{Meshes and broadside $E_x$ and $E_y$ fields of the survey domain in the left, middle, and right column, respectively, for the rectilinear, octree, and tetrahedral meshes in the top, middle, and bottom row, respectively (3$\times$ vertical exaggeration).}
  \label{fig:results-marlim_survey}
\end{figure*}

Figure~\ref{fig:results-marlim_survey} shows a zoom-in to the survey domain, the domain of interest, which shows how the different codes discretized the zones of interest.
The most striking point of these mesh comparison plots is that, while the overall fields are similar and the responses at receiver locations agree well, the fields can vary quite a lot in different parts of the model. Two kilometres below the seafloor, as an example, there are already significant changes visible, purely due to meshing. This is an important insight that has to be taken into account when interpreting modelling results and the feasibility for inversions.

The required runtime and memory to compute the shown MR3D responses for the four codes are listed in Table~\ref{tbl:comp-marlim}. The memory requirement varies from $0.5-230\,$GiB, and the runtime from roughly 7\,min to 20\,min, where the codes were run on very different machines from servers to supercomputers (see Table \ref{tbl:machines}) and use between 1 and 96 processes in parallel.
\begin{table*}
\begin{minipage}{10cm}
  \centering
  \caption{Comparison of number of processes, runtime, and memory, as well as the degree of freedom of the discretization used by the different codes for the MR3D model.}
\label{tbl:comp-marlim}
  \begin{tabular}{lrS[table-format=6.0]S[table-format=4.1]S[table-format=8.0]}
  \toprule
  Code & \#Procs & {Runtime (s)} & {Memory (GiB)}   & {\#dof} \\
  \midrule
  \custem & 64 &  872 & 230.1 & 1918106 \\ 
  \emg3d  &  1 & 1246 &   0.5 & 5998992 \\
  \petgem & 96 &  524 & 175.4 & 1918106 \\
  \simpeg &  4 &  422 &  12.8 &  720146 \\ 
  \bottomrule
\end{tabular}
\end{minipage}
\end{table*}

\begin{figure*}
  \centering
  \includegraphics[width=.9\fwidth]{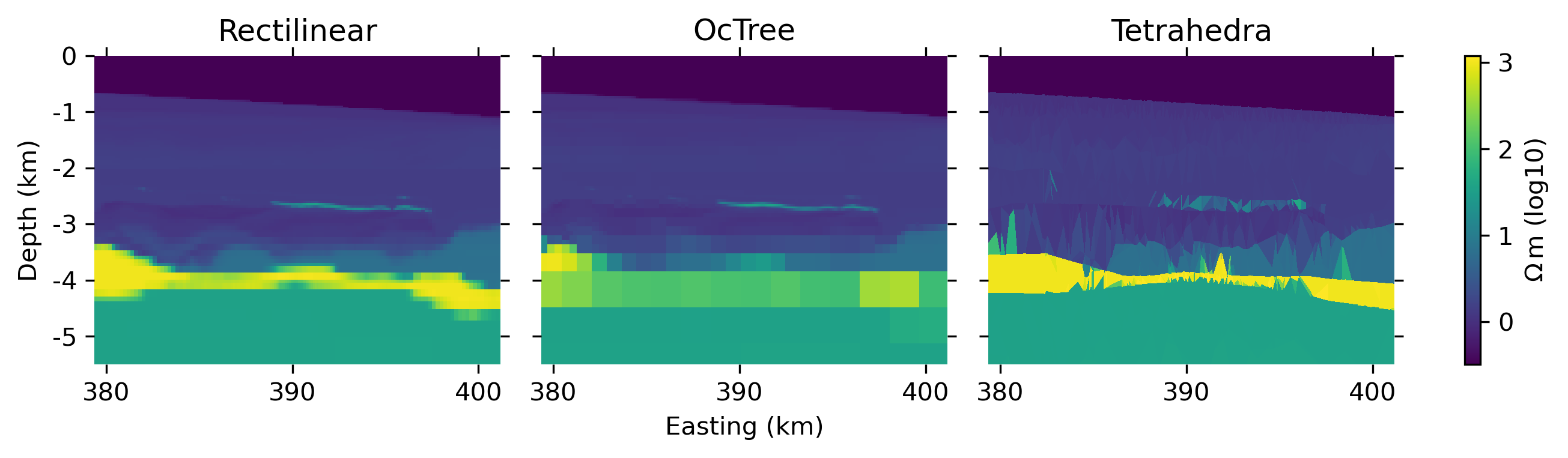}
  \caption{The resistivity model as obtained from the three different discretizations using rectilinear, octree, and tetrahedral meshes (3$\times$ vertical exaggeration).}
  \label{fig:results-marlim_allmodels}
\end{figure*}

The actual discretized resistivity model of the three mesh types are shown in Figure~\ref{fig:results-marlim_allmodels}.
The models where obtained using volume averaging (\emg3d, \simpeg) and linear interpolation (\custem, \petgem), in both cases on $\log_{10}(\rho)$ values. The actual representation of the original resistivities in the tetrahedra mesh is not as smooth as in the other meshes since this would require significantly smaller tetrahedra and more dof on top of the inclusion of subsurface layer constraints. However, note that the combination of slicing through tetrahedra and the vertical exaggeration leads to a very distorted view of the tetrahedral mesh which is barely representative for the overall averaged approximation of resistivities. The actual responses at the receivers (field interpolation) are obtained by spline interpolation (\emg3d), linear interpolation (\simpeg), and through the second-order basis functions (\custem, \petgem). It is worth mentioning as a final note that we used frequency independent meshes in this study. Creating frequency dependent meshes could potentially improve both the result and the runtime.

\section{Discussion}

Today we are in the fortunate situation where we have several open-source codes available to model CSEM data for arbitrarily sized complex 3D models. It is therefore comparably easy for anyone to run a simulation and obtain reasonable-looking, robust and precise results, something that was impossible a few years ago. However, just because one obtains a good-looking result does not say anything about its accuracy. Though we may be able to confirm the accuracy of codes in some simple scenarios, validating the performance of 3D CSEM codes for complex problems is only possible by cross-validating multiple solutions. This, plus the necessity for more reproducible modelling results, are the core motivations for this study.

Overall, we observed an excellent match between the solutions. The amplitudes and phases of the layered model have mostly a relative error of less than 1--2\,\%. The cross-validation of the results for the block model yields a similar picture, with normalised differences within a few percents. For the large, complex MR3D model it is a bit different. First it can be noted that the misfits are higher for higher frequencies. For the lowest frequency, 0.125\,Hz, the normalised difference is a few percent for inline and broadside $E_x$ components, and up to 10\,\% for the broadside $E_y$ component. For higher frequencies these numbers increase to the order of $\pm10\,$\%. The reason for the higher misfit for MR3D than for the layered or block model lies mostly in the discretization, boundary effects, and interpolation strategies for the model, the source distribution on the mesh, and to obtain the response at receiver locations. While one might argue that 5--10\,\% NRMSD is a lot, we argue that this is probably as good as it gets, given that the code with which the responses we compare to were computed is not open-source. This has been shown within the process of this study. Our first attempts were based on the originally published, fine MR3D model. This model is very detailed, but its extent is too small in the x- and y-directions for CSEM computations. In order to represent the fine-scale model on a mesh suitable for computation, each of our codes did its own upscaling and extrapolation. However, we were unsuccessful in our attempt to reproduce the published results.  The reason is that there are many ways how you can extrapolate a model, and without knowing the details of how MR3D represents the model on the computational mesh, we could not re-produce this step. We reached out to the authors of MR3D, and they kindly agreed to also publish the upscaled and extended computational model. It is only with this model that we were able to obtain comparable results.

One of the key insights we gained from this study is that creating appropriate meshes is a difficult and time-consuming task. While there now exist several open-source 3D CSEM codes the capability for automatic meshing lacks behind and currently remains a largely manual task requiring experience in the field. It is important to state that the chosen models favour structured meshes. The first two examples can easily and accurately be represented by regular meshes. Using, for instance, a dipping layer instead of three resistive cubes would turn it around and make the model more favourable for unstructured meshes. MR3D would be an ideal scenario for unstructured meshes, as they are best suited to represent the irregular geometry provided by the lithological horizons. Since the corresponding resistivity model, however, was defined on a regular grid, the FE codes were forced to approximate the comparatively fine regular discretization by an unstructured one and interpolate the resistivity data for being able to apply the FE codes to this problem. We are completely aware about the ineptness of this re-approximation procedure, unless it served for the cross-validation purposes. Both cases show that automatic and adaptive meshing capabilities are needed, which will increase the flexibility of 3D CSEM modelling \citep{GJI.11.Schwarzbach,GJI.16.Key,GJI.19.CastilloReyes}. Such meshing codes should also take into account the physics of CSEM computation with its diffusive behaviour. It is also important to state that the unstructured mesh used for the MR3D model incorporates the entire domain and could be used for all source and receiver positions, whereas the rectilinear and octree meshes were designed for the shown receiver and corresponding source lines. Also, creating FV meshes is much simpler than creating FE meshes. However, having a FE mesh yields a lot of flexibility with regards to interpolation and local refinement, making it more powerful once created.

Whereas proper discretization is the main driver for accurate results, the used solver is the main discriminator in terms of runtime and memory. Both iterative and direct solvers are used in this study, however, the chosen models clearly favour iterative solvers. The biggest advantage of iterative solvers are the comparatively small memory requirements, even more so if the iterative solver is matrix free. Direct solvers exhibit their strongest advantage, besides stability and robustness, in terms of computation times if responses of multiple CSEM transmitters need to be computed in the same computational domain. As the system matrix factorisation requires 98--99\,\% of the solution time, computations for additional sources come at almost no cost for direct solvers, whereas it would come at the same cost as the first source for iterative solvers. Independently of the solver there is a trade-off between high accuracy and runtime. For example, the octree mesh made for a very fast computation in the MR3D example thanks to the low number of dof, even though a direct solver was used. But the octree mesh has, due to its 2:1 aspect ratio, some limitations on the accuracy, why it was not used in the first two examples. Having discussed the discretization and the solver it is important to state that the reported runtime and memory is just one of the aspects, and we would like to emphasise that neither was at the core of this comparison, but the validation of the results. As such, no special efforts were undertaken to minimise either, as this is an entire different task.

A positive outcome of a collaboration between different projects such as this is that it brings the realm as a whole further, which should be motivation enough for further collaborations. Within \simpeg, this work has motivated feature development including averaging strategies to map physical properties on a fine mesh to a coarser mesh for computation and new examples to be included in the documentation for designing octree meshes. Furthermore, as a part of a broader development objective of inter-operating with other forward-simulation engines, connecting \emg3d and \simpeg provided a motivating use-case for the latest refactor and release of \simpeg 0.14.0. Within \emg3d, this work pushed a lot of the meshing functionality, and implementation of I/O utilities for different file formats. Within \custem, cell-wise resistivity interpolation was added for this work, multi-layer subsurface topography was the first time applied as well as first time reciprocity modelling, and also improved mesh design (resolution, refinement, etc.). Within \petgem, this work has promoted the inclusion of routines for magnetic field computation and the implementation of continuous integration functions. It has also improved its test suite and work-flow for the generation of adapted meshes based on a semi-automatic approach that reduces user intervention.

The shown examples consider the geophysical problem of marine CSEM with resistive bodies in the frequency domain. Marlim R3D is a rare example of an open-source resistivity model that also comes with simulated CSEM data. The spectrum of geophysical EM methods is much wider than marine CSEM. We hope to see in the future more open-source models with accompanying data and similar comparisons for other cases, such as land CSEM with strong topography; airborne and mixed air-ground surveys; looking for conductive bodies; time domain; and models based on horizons instead of blocks which pose a challenge for structured grids and favour unstructured meshes. An ideal realistic model should be defined as a function \texttt{resistivity(x, y, z)} which returns the resistivity for any location in the domain of interest, preferably with the horizons of the major formations. This would allow any meshing strategy to extract its optimal resistivity model.

It was a fruitful path to get four different CSEM codes to work together, and it included a steep learning curve for all involved parties. Our comparison and our results are far from perfect. The chosen models favour regular grids, and the chosen surveys favour iterative solvers. We only consider one particular CSEM case, the shallow marine setting looking for a resistor. Our comparison raised probably more questions regarding the accuracy of 3D CSEM modelling than it answered. We consider this as an important, initial step which we believe yields already many insights to the reader. We have shown similarities and differences in FV and FE codes and our results show that ideally the CSEM responses of a complex model should be computed with various codes, as there is not a single code that will provide very accurate results. Focusing on reproducibility and open-source software and data forced us to show the misfit between the codes brutally honest, there is no hiding. We think this should be done more often. We do not claim that our codes are superior to other codes in any way. It is our codes' current status, and we hope that we can build on this and that many similar comparisons will follow.

\section{Conclusions}

We compare the computed fields and underlying meshes of four different open-source 3D CSEM codes by means of three increasingly complex resistivity models. A layered marine, anisotropic model is used as first example to verify our results with semi-analytical responses. All codes show excellent accuracy with a relative error in the order of a few percents. Three resistive blocks are added to the layered model to increase complexity in the second example. The cross-validation of the outcomes of the different codes shows a normalised difference of a few percent, hence in a similar range as the relative error in the layered model. The third example is Marlim R3D, a realistic, complex, marine resistivity model. The MR3D model and corresponding CSEM responses are open-source, allowing an independent validation of our results with results from another (closed-source) code. In this scenario the normalised difference is in the order of 5--10\,\%, where the main source of error can be attributed to differences in discretization and handling of the bathymetry with corresponding advanced interpolations of the electromagnetic fields. The required runtime and memory consumption is primarily controlled by the used solver. The accuracy, on the other hand, depends to a large degree on the mesh. A proper discretization is therefore key for an accurate result, which makes this step the most time-consuming task, not the actual computation of the responses itself. Validating the correctness of a 3D code is a difficult task, and it is essential to have easily accessible benchmark models with reliable and reproducible solutions. Our study is one example of a collaboration facilitated by open practices including sharing of code and data. We hope that these results may be useful for the entire CSEM community at large, and we invite and encourage the community to make more code, modelling scripts, and results publicly available.

\section{Data}
The resulting CSEM responses of the four codes for the three models, as well as all files to rerun the different models with the four codes and reproduce the shown results are available at Zenodo (\href{https://doi.org/10.5281/zenodo.4535602}{10.5281/zenodo.4535602}).

\begin{acknowledgments}

We would like to thank Bane Sullivan for plotting of tetrahedra meshes in \texttt{matplotlib} through \texttt{PyVista}, Seogi Kang for the input and octree mesh design for the \simpeg layer and block models, and Joseph Capriotti for the octree mesh implementation and volume averaging in \simpeg. We would also like to thank Paulo Menezes for the help and explanations with regards to the Marlim R3D model and corresponding CSEM data, and for making their actual computation model available under an open-source license. We would further like to thank the editor Ute Weckmann and assistant editor Fern Story as well as the reviewers Colin Farquharson and Rune Mittet for many helpful comments, which improved this manuscript considerably.

The work of D.W. was conducted within the Gitaro.JIM project funded through MarTERA, a \emph{European Union's Horizon 2020} research and innovation programme, grant agreement N$^\circ$~728053.

The development of \custem by R.R. as part of the DESMEX/DESMEX II projects was funded by the Germany Ministry for Education and Research (BMBF) in the framework of the research and development program Fona-r4 under grants 033R130D/033R130DN.

The work of O.C-R. has received funding from the \emph{European Union's Horizon 2020 programme} under the \emph{Marie Sklodowska-Curie} grant agreement N$^\circ$ 777778. Further, the development of \petgem has received funding from the \emph{European Union's Horizon 2020 programme}, grant agreement N$^\circ$~828947, and from the Mexican Department of Energy, CONACYT-SENER Hidrocarburos grant agreement N$^\circ$ B-S-69926. Furthermore, O.C-R. has been 65\% cofinanced by the European Regional Development Fund (ERDF) through the Interreg V-A Spain-France-Andorra program (POCTEFA2014-2020). POCTEFA aims to reinforce the economic and social integration of the French-Spanish-Andorran border. Its support is focused on developing economic, social and environmental cross-border activities through joint strategies favouring sustainable territorial development.

The work of L.H. received funding from the National Science Foundation EarthCube program under award 1928406.
\end{acknowledgments}

\bibliographystyle{gji}
\bibliography{Refs}

\label{lastpage}

\end{document}